\documentclass[aps,prd,eqsecnum,showpacs,superscriptaddress,
twocolumn,nofootinbib,floatfix,preprintnumbers,altaffilletter]{revtex4}
\usepackage{graphicx}
\usepackage{hyperref}
\usepackage{amssymb}
\usepackage{amsmath}
\usepackage{longtable}
\usepackage{rotating}
\usepackage{color}
\usepackage{fancyhdr}

\def\lesssim{\mathrel{\hbox{\rlap{\hbox{\lower4pt\hbox{$\sim$}}}\hbox{$<$}}}}
\def\gtrsim{\mathrel{\hbox{\rlap{\hbox{\lower4pt\hbox{$\sim$}}}\hbox{$>$}}}}
\def\cos{\rm cos}
\def\sin{\rm sin}
\newcommand{\be}{\begin{equation}}
\newcommand{\ee}{\end{equation}}
\newcommand{\bea}{\begin{eqnarray}}
\newcommand{\eea}{\end{eqnarray}}
\newcommand{\bdm}{\begin{displaymath}}
\newcommand{\edm}{\end{displaymath}}

\begin{document}
\title{The generalized ${\mathcal F}$-statistic: multiple detectors and multiple GW pulsars}
\author{Curt Cutler}
\author{Bernard F. Schutz}
\affiliation{Max-Planck-Institut f\"ur Gravitationsphysik (Albert-Einstein-Institut), Am M\"uhlenberg 1, 14476 Golm, Germany}

\date{\today}

\begin{abstract}
The ${\mathcal F}$-statistic, derived by Jaranowski, Krolak \& Schutz (1998), is the
optimal (frequentist) statistic for the detection of nearly periodic
gravitational waves from known neutron stars, in the presence
of stationary, Gaussian detector noise.  The ${\mathcal F}$-statistic was originally derived for the case of a 
single detector, whose noise spectral density was assumed constant in time, 
and for a single known neutron star.
Here we show how the ${\mathcal F}$-statistic can be straightforwardly generalized 
to the cases of 1) a network of detectors with time-varying
noise curves, and 2) a collection of known sources (e.g., all
known millisecond pulsars within some fixed distance). 
Fortunately, all the important ingredients that go into our
generalized ${\mathcal F}$-statistics are {\it already} calculated in
the single-source/single-detector searches that are currently 
implemented, e.g., in the LIGO Software Library, so implementation of
optimal multi-detector, multi-source searches should require negligible
additional cost in computational power or software development.
This paper also includes an analysis of the likely efficacy of  
a collection-type search, and derives criteria for deciding which 
candidate sources should be included in a collection, if one
is trying to maximize the detectability of the whole. In particular 
we show that for sources distributed uniformly in a thin disk, 
the strongest source in the collection should have signal-to-noise-squared
$\sim 5$ times larger than weakest source, for an optimized
collection. We show that gravitational waves from 
collection of the few brightest (in gravitational waves) neutron stars
could perhaps be detected before the single brightest source, but that
this is far from guaranteed. Once gravitational waves from the
few brightest neutron stars have been
discovered, grouping more distant (individually undetectable) pulsars 
into collections, and then searching for those collections, 
should be an effective way of measuring the average gravitational-wave
strengths of those more distant pulsars.

\end{abstract}

\maketitle 

\section{Introduction}
\label{Sec:Int}

The ${\mathcal F}$-statistic, as first derived by
Jaranowski, Krolak \& Schutz~\cite{jks} (hereinafter referred to 
as JKS), is the optimal frequentist statistic for the 
detection of nearly periodic gravitational waves (GWs) from a known 
neutron star. In the original JKS version,   
the ${\mathcal F}$-statistic was derived only for the  
case of a single GW detector (which was assumed to have stationary noise 
characteristics) and a single known neutron star 
(assumed to be emitting GWs at the neutron star's rotation frequency and/or
at twice its rotation freuqency).
Here we show how the ${\mathcal F}$-statistic can be generalized in a 
straightforward manner 
to the cases of 1) a network of detectors with time-varying
noise curves, and 2) an entire collection of known sources. 
Fortunately, all the important ingredients that go into the 
generalized ${\mathcal F}$-statistic are {\it already} calculated in
the single-detector/single-source searches that are currently 
implemented, e.g., in the LIGO Software Library~\cite{lscsoft}, 
so implementation of
optimal multi-detector and/or multi-source searches should require negligible
additional cost in software development and computation.

We note that the problem of optimally combining data from different 
detectors has already been solved for several types of GW searches.
For the case of inspiralling binaries, we refer the reader to
Bose, Pai \& Dhurandhar~\cite{Bose} and to
Finn~\cite{Finn}; for the case of GW bursts, to 
Sylvestre~\cite{Sylvestre}; and for the case of LISA observations of
galactic, stellar-mass binaries, to 
both Krolak et al.~\cite{Krolak_Tinto_Vallisneri} and 
Rogan \& Bose~\cite{Rogan_Bose}. (LISA can effectively be treated 
as a network of three independent GW detectors.) Our analysis  
in Sec.~III is especially similar to that of 
Krolak et al.~\cite{Krolak_Tinto_Vallisneri} 
and Rogan \& Bose~\cite{Rogan_Bose}, since formally the sources considered
there are equivalent to GW pulsars. Like GW pulsars, the stellar-mass
binaries visible to LISA are effectively monochromatic sources that can
be characterized by four amplitude parameters, in addition to the GW frequency and the  
two angles specifying the source position on the sky.

The basic idea of somehow combining the signals 
from many individually-undetectable sources or events, in hopes of finding 
a statistical excess, is also hardly a new one.
In GW astronomy, a good example is the suggestion of 
looking for GW bursts associated with gamma-ray bursts by 
cross-correlating the outputs of LIGO's L1 and H1 detectors 
over short time windows  coincident with hundreds of observed gamma-ray 
bursts~\cite{Finn_Mohanty_Romano}. But our application of 
this idea to the population of known millisecond pulsars appears to be new.
We investigate when this strategy is likely to be effective and 
derive useful criteria for deciding how many and which sources 
should be included in the collection, in order to maximize the
detectability of that group.  

The plan of this paper is as follows. In Sec.\ \ref{SecII} we briefly 
establish notation; we generally try to align our notation with that 
of JKS, to ease comparison with their work.
In Sec.\ \ref{network} we derive the ${\mathcal F}$-statistic for a network of 
$N$ detectors and a single source. This multi-detector ${\mathcal F}$-statistic follows
a $\chi^2$ distribution with $4$ degrees of freedom, exactly as with
the single-detector version. We consider the general case where the
detectors have correlated noises, but of course our expressions simplify
in the case where noises from different detectors are uncorrelated. 
As a bonus, our results immediately show how to 
appropriately time-weight the data in the (realistic) case that the detector
noise spectra are slowly time-varying. (The appropriate time-weighting for
a single detector has already been 
derived by Itoh et al.~\cite{itoh} 
and is implemented in the LIGO Software Library. 
We give an independent derivation, since here it follows trivially.)

In Sec.\ \ref{collection} we extend the ${\mathcal F}$-statistic to the case of a collection of
known sources.  If there are $M$ known sources 
(each emitting at a single, known
frequency), then the correct ${\mathcal F}$-statistic for the entire collection 
follows a $\chi^2$ distribution with $4M$ degrees of freedom.
(This is true for both the single-detector case and for an $N$-detector 
network.)
The most interesting target population is clearly (some subset of) the
known millisecond pulsars. We consider two particularly interesting 
cases: 1) a collection of the
few very brightest GW pulsars, and 2) a larger collection of more distant
GW pulsars.
We investigate the expected gains from both these types of multi-source
searches, under the reasonable assumption that there exists some 
population of GW pulsars
that is uniformly spread thoughout the Galactic disk. 
As a further illustration of multi-source searches,  
we estimate the sensitivity of the LIGO network to the collection 
consisting of the five ``most promising'' millisecond pulsars, assuming they 
all have the same ellipticity.  
Our conclusions are summarized in Sec.\ \ref{conclusion}.

\section{Notation}
\label{SecII}

Let us consider an N-detector network, with output $x^{\alpha}(t)$, 
$\alpha = 1,...,N$. (In cases where a single instrument outputs
$k$ independent data streams--e.g., a spherical bar detector that 
encodes for two GW polarizations in five data streams--we
regard these formally as the outputs of $k$ different detectors.)
For simplicity, we begin by assuming that the detector noise 
is both stationary and Gaussian. 
We allow, however, for the possibility that the noises are correlated.
Then we have
\begin{equation}
\label{p1}
<{\tilde n}^{\alpha}(f) \, {\tilde n}^{\beta}(f^\prime)^*> = {1 \over 2}
\delta(f - f^\prime) S_h^{\alpha \beta}(f) \, .
\end{equation}
Here tildes denote Fourier transforms, according to the convention that 
\be
{\tilde x}(f) = \int_{-\infty}^{\infty} e^{2 \pi i f t} x(t) dt \, ,
\ee
and ``$\left< \cdots \right>$'' denotes ``expectation value''.
We note that $S_h^{\alpha \beta}(f)$ is the {\it single-sided} noise 
spectral density, which is also the convention followed in JKS.
(If we were using the {\it double-sided} convention, the factor ${1 \over 2}$ 
on the rhs of in Eq.~(\ref{p1}) would be replaced by 1.)

The Gaussian random process ${\bf n}(t)$ determines a natural inner
product $\left( \ldots | \ldots \right)$ on the space of 
functions ${\bf x}(t)$~\cite{finn_92}:
\FL
\begin{equation}
\label{product_def}
\left( {\bf x} \, | \, {\bf y} \right) \equiv 4 \, \Re \int_0^\infty df
\,\, {\tilde x}^{\alpha}(f)^* \left[ S_h^{-1}(f) \right]_{\alpha \beta}
{\tilde y}^{\beta}(f),
\end{equation}
where $\left[ S_h^{-1}(f) \right]_{\alpha \beta} S_h^{\beta \gamma}(f) 
= \delta_{\alpha}^{\gamma}$ and 
where $\Re$ means ``the real part of''.  
Here and below, to reduce index clutter, we sometimes represent a 
signal vector, having one
component for each detector, by simply using boldface
without an index; e.g., ${\bf x}(t)$ instead of  $x^{\alpha}(t)$.
The inner product Eq.~(\ref{product_def}) is such that
the probability distribution function (pdf) for the noise  ${\bf n}(t)$
takes the form
\begin{equation}
\label{noise_distribution}
{\rm pdf}[{\bf n}] \,= \, {\cal N} e^{- \left( {\bf n} | {\bf n}\right) / 2 },
\end{equation}
\noindent where $\cal N$ is a normalization constant.
It follows that the expectation value 
of the product $\left( {\bf x} \, | \, {\bf n} \right) \left( {\bf y} \, | \, {\bf n} \right)$, over many realizations of the noise, is simply given by
\be\label{expectation}
\left< \, \left( {\bf x} \, | \, {\bf n} \right) \left( {\bf y} \, | \, {\bf n} \right) \, \right> = \left( {\bf x} \, | \, {\bf y} \right) \, .
\ee

\section{${\mathcal F}$-statistic for a Detector Network}
\label{network}

Given gravitational-wave data 
from a single detector, 
the ${\mathcal F}$-statistic developed by 
JKS is the optimal frequentist statistic for the detection 
of GWs from a single known NS
in that single data stream. This section answers the question: if we
have data from a network of detectors (possibly including bars as well
as interferometers)
how  does one combine the different
data streams to produce the optimal detection statistic
for the entire network?~\footnote{JKS briefly consider 
this question and sketch a claimed answer, in \S 4 of their paper~\cite{jks}, 
but their answer is quite wrong. In particular, they claim that the 
appropriate ${\mathcal F}$-statistic for a network with  N detectors follows 
a $\chi^2$ distribution with 
$4N$ degrees of freedom, but we shall see below that the right
number of degrees of freedom is just $4$--the same as
for the single-detector case. 
This is because there are still just four unknowns in
the problem: the amplitude and phase of each of the two GW polarizations.}

\subsection{${\mathcal F}$-statistic for Single Source and Multiple Detectors, all 
with Time-Invariant Noise Curves}
\label{multiple}

Consider the search for nearly periodic GWs from a single
source with known position and known (possibly time-varying) 
frequency, e.g., PSR 1937+21. 
The GW signal is characterized by
four unknowns: an overall amplitude $A_0$ (equivalent to the combination
$h_0 {\rm sin}\zeta {\rm sin}^2\theta$ in the notation of JKS), two angles
$\iota$ and $\psi$ that characterize the waves' polarization (equivalent 
to determining the direction of the NS's spin axis), and 
an overall phase $\Phi_0$.  The GW signal $h^{\alpha}(t)$ depends nonlinearly
on $\iota, \psi, \Phi_0$, but, crucially, one can make a simple change of 
variables--to $\big(\lambda^1, \lambda^2, \lambda^3, \lambda^4 \big)$--such
that dependence of $h^{\alpha}(t)$ is linear in these new variables:
\be 
h^{\alpha}(t) = \sum_{a=1}^4 \lambda^a h_a^{\alpha}(t)
\ee
\noindent where the four basis waveforms $h_a^{\alpha}(t)$ are
defined by
\FL
\bea\label{hdef2}
&&h_1^{\alpha}(t) = F_+^{\alpha}(t) \cos \Phi^{\alpha}(t) , \   
h_2^{\alpha}(t) = F_{\times}^{\alpha}(t) \cos \Phi^{\alpha}(t) , \ \ \ \ \ \ \ \ \ \nonumber \\
&&h_3^{\alpha}(t) =  F_+^{\alpha}(t) \sin \Phi^{\alpha}(t) , \
h_4^{\alpha}(t) = F_{\times}^{\alpha}(t) \sin \Phi^{\alpha}(t)  \, .\ \ \ \ \ \ \ \ \ 
\eea
Here $\Phi(t)$ is the waveform phase at the detector:
\be \label{Phase}
\Phi(t) \approx 2\pi \int^{t} f_{gw}(t') dt'  \, ,
\ee
where $f_{gw}(t')$ is the measured GW frequency at the detector at
time $t'$.  The measured frequency includes the Doppler effect from the
detector's motion relative to the source, as well as Einstein and Shapiro
delays associated with the Earth's orbit around the Sun. 
When the GW pulsar is in 
a binary, then $f_{gw}(t')$ also includes the Roemer, Einstein, and
Shapiro delays associated with that binary orbit.  We emphasize that 
the known-pulsar searches described here do {\it not} require the
GW pulsar be isolated, but just that there exist an accurate
timing model for the emitted waves. 
The $F_+^{\alpha}(t)$ and $  F_{\times}^{\alpha}(t)$ terms in 
Eq.~(\ref{hdef2}) are the 
beam-pattern functions giving the response of the $\alpha^{th}$ detector 
to the $+$ and $\times$ polarizations, respectively. 
We note that the exact form of $F_+^{\alpha}(t)$ and $F_{\times}^{\alpha}(t)$ 
depends on one's convention for decomposing the waveform into ``plus'' and
``cross'' polarizations; a one-parameter family of choices is possible,
corresponding to the freedom to rotate the axes around the line of sight.
JKS follow the conventions of 
Bonazzola \& Gourgoulhon~\cite{Bonazzola}.

A further word on our index notation: 
as above, we use Greek 
indices from the beginning of the alphabet ($\alpha, \beta, \gamma$) 
to indicate the various detectors in the network;
we use Latin letters from the beginning of the alphabet ($a,b,c$)
to indicate the four independent waveform components from a single NS
(emitting at a single frequency);  and we use
Latin letters from the middle of the alphabet ($i,j,k$)
to label different NSs.
As above, we sometimes remove the
Greek index and instead represent the vector in boldface: 
${\bf h}_a(t)$ instead of $h^{\alpha}_a(t)$.
Finally, we use the capital Latin letter ``J'' to  
label different time intervals (always intervals
over which the noise spectral density $S^{\alpha \beta}_h(f)$ can be safely
approximated as constant);

Next we define the $4\times 4$ matrix $\Gamma_{ab}$ by 
\be
\Gamma_{ab} \equiv \big(\frac{\partial {\bf h}}{\partial \lambda^a}\, | \, \frac{\partial {\bf h}}{\partial \lambda^b}\big) 
= \big({\bf h_a} \, | \, {\bf h_b}\big)  \, .
\ee
Because both the observation time and $1$ day (the timescale on which
the $F_{+,{\times}}^{\alpha}(t)$ vary) are vastly larger than 
the period of the sought-for GWs (typically $10^{-2}-10^{-3}$ s), 
we can 
replace $\cos^2\Phi(t)$,  $\sin^2\Phi(t)$, and $\cos\Phi(t) \sin \Phi(t)$ 
by their time-averages: $\cos^2\Phi(t),\sin^2\Phi(t) \rightarrow \frac{1}{2}$, while $\cos\Phi(t) \sin \Phi(t) \rightarrow 0$. Then we have
\bea
\Gamma_{11} &\approx& \sum_{\alpha, \beta} \big(S_h^{-1}(f_{gw})
\big)_{\alpha \beta}
\int{F_+^{\alpha}(t)F_+^{\beta}(t)\, dt} \nonumber \\
\Gamma_{12} &\approx& \sum_{\alpha, \beta} \big(S_h^{-1}(f_{gw})\big)_{\alpha \beta}
\int{F_+^{\alpha}(t)F_{\times}^{\beta}(t) \, dt} \nonumber \\
\Gamma_{22} &\approx& \sum_{\alpha, \beta} \big(S_h^{-1}(f_{gw})\big)_{\alpha \beta}
\int{F_{\times}^{\alpha}(t)F_{\times}^{\beta}(t) \, dt} \, ;
\eea
\noindent additionally, $\Gamma_{33} \approx \Gamma_{11}$, 
 $\Gamma_{34} \approx \Gamma_{12}$, $\Gamma_{44} \approx \Gamma_{22}$, and
 $\Gamma_{13} \approx \Gamma_{14} \approx \Gamma_{23} \approx \Gamma_{24} \approx 0$.

The best-fit values of $\lambda^a$ satisfy 
\be
{\frac{\partial }{\partial \lambda^a } } ({\bf x} - \sum_b \lambda^b {\bf h_b}\,  | \, x - \sum_c \lambda^c {\bf h_c}) = 0
\ee
\noindent implying 
\be
\lambda^a = \sum_b \, (\Gamma^{-1})^{ab}({\bf x} \, | \, {\bf{h_b} }) \, ,
\ee
\noindent
and our optimal statistic $2{\mathcal F}$ is then just twice the log 
of the likelihood ratio:
\bea\label{F1} 
2{\mathcal F} 
&=& ({\bf x} | {\bf x} ) - ({\bf x} - \sum_b \lambda^b {\bf h_b} \ \, | \, \ {\bf x} - \sum_c \lambda^c {\bf h_c}) \nonumber \\
&=& \sum_{a,d} \, (\Gamma^{-1})^{ad}({\bf x}|{\bf h_a}) ({\bf x}|{\bf h_d}) \, .\label{F2}
\eea
Therefore using $2{\mathcal F}$ as one's detection statistic satisfies the
Neyman-Pearson criterion for an optimum test: it minimizes the 
false dismissal (FD) rate for any given false alarm (FA) rate. 

Writing ${\bf x} = {\bf n} + {\bf h}$, and plugging into Eq.~(\ref{F2}), we
find 
\be 
\left< 2{\mathcal F} \right> = 4 + ({\bf h}\  |\ {\bf h}) \, ,
\ee
where we have used Eq.~(\ref{expectation}) and the fact that \linebreak
$ <({\bf h}\  |\ {\bf n})> \ = \  0 $.  More generally, it is easy to show
that $y \equiv 2{\mathcal F}$ follows a $\chi^2$ distribution with
$4$ degrees of freedom (d.o.f) and non-centrality parameter $\rho^2 \equiv 
({\bf h} | {\bf h})$:
\be\label{py}
P(y) = \chi^2(y |4; \rho^2) \, .
\ee

\subsection{${\mathcal F}$-statistic for a Single Source and Multiple Detectors with
Time-Varying Noise Curves}
It is trivial to generalize the above results to a network of detectors with 
time-varying noise curves. Divide the total observation time into segments that
are short enough that all noise correlation functions $S_h^{\alpha \beta}$ can be 
approximated as constant during each segment. (We assume the segments are still much
longer than the GW period.) Let there be $p$ such segments in all.
Let the beginning and end points of these time intervals be $(t_0, 
t_1, \cdots , t_p)$. (In this scheme, we can formally represent gaps in the output of
one or more detectors by intervals where some of the components  $S_h^{\alpha \beta}$
go to infinity.)
While our signals come from N detectors with time-varying noise curves, we can formally
regard them as coming from $p\, N$ detectors, each with stationary noise (but such
that only $N$ detectors are turned ``on'' at any instant; when $N$ more turn on, the
previous $N$ turn off). But we know how to construct the ${\mathcal F}$-statistic for 
$p\, N$ detectors with stationary noise characteristics, from the previous
subsection. (Nothing in that subsection required all the detectors to 
be ``on'' simultaneously.) The noise spectral density coefficients 
$S_h^{\alpha \beta}(f)$ are now labelled by time interval $J$: 
$S_{h,J}^{\alpha \beta}(f)$. Then $\Gamma_{11}$ becomes
\FL
\bea
\Gamma_{11} &\approx& \sum_{\alpha, \beta =1}^{N} \sum_{J=1}^{p}\big(S^{-1}_{h,J}(f_{gw})\big)_{\alpha \beta}
\int_{t_{J-1}}^{t_J}{F_+^{\alpha}(t)F_+^{\beta}(t) \, dt} \ \ \nonumber \\
&\rightarrow& 
\sum_{\alpha, \beta =1}^{N} 
\int_{t_0}^{t_p}{ \big(S^{-1}_h(f_{gw},t)\big)_{\alpha \beta} 
F_+^{\alpha}(t)F_+^{\beta}(t) \, dt}\ \ \ 
\eea
\noindent where we have made the notational shift 
$S_{h,J}^{\alpha \beta}(f_{gw}(t)) \rightarrow 
S_h^{\alpha \beta}(f_{gw}(t), t)$; i.e., we have replaced the discrete label
``J'' by the continuous label ``t''.  (In practice, 
the noise spectral density at any instant is estimated from the
data itself, e.g., by use of a running mean.)

Similarly,
\bea
\Gamma_{12} &\approx & 
\sum_{\alpha, \beta =1}^{N} 
\int_{t_0}^{t_p}{ \big(S^{-1}_h(f_{gw}(t),t)\big)_{\alpha \beta} 
F_+^{\alpha}(t)F_{\times}^{\beta}(t) \,dt} \ \ \ \ \ \nonumber \\
\Gamma_{22} &\approx & 
\sum_{\alpha, \beta =1}^{N} 
\int_{t_0}^{t_p}{ \big(S^{-1}_h(f_{gw}(t),t)\big)_{\alpha \beta} 
F_{\times}^{\alpha}(t)F_{\times}^{\beta}(t) \, dt}\, . \ \ \ \ \ \ \ 
\eea
\noindent and again $\Gamma_{33} \approx \Gamma_{11}$, 
 $\Gamma_{34} \approx \Gamma_{12}$, $\Gamma_{44} \approx \Gamma_{22}$, while
 $\Gamma_{13} \approx \Gamma_{14} \approx \Gamma_{23} \approx \Gamma_{24} \approx 0$.

If we define
\be 
({\bf x} \, | \, {\bf h_a}) \equiv \sum_{\alpha, \beta}\,
\int_{t_0}^{t_p}{ \big(S^{-1}_h(f_{gw}(t),t)\big)_{\alpha \beta} 
x^{\alpha}(t) h_a^{\beta}(t)  dt} \, ,
\ee
then Eq.~(\ref{F2}) remains the correct expression for $2{\mathcal F}$, and
Eq.~(\ref{py}) remains its correct distribution function, 
with $y \equiv 2{\mathcal F}$ and $\rho^2 = ({\bf h} | {\bf h})$.
That is, given our notation, the expression for the multi-detector
${\mathcal F}$-statistic is the same as for the single-detector case.

\subsection{${\mathcal F}$-stastic for a Single Source and N Detectors with
Uncorrelated Noises}

The expressions simplify somewhat in the (common) case where
the noises from different detectors are uncorrelated: 
$S_h^{\alpha \beta}(f,t)  = 
S_h^{\alpha}(f,t) \delta^{\alpha \beta}$. Then the inner product
$({\bf x}\, | \,  {\bf y})$ is given by
\be
({\bf x}\, | \,  {\bf y}) \equiv  2 \sum_{\alpha} \int_{t_0}^{t_p}
\frac{x^{\alpha}(t)y^{\alpha}(t)\,dt}{S^{\alpha}_h(f_{gw}(t),t)}  \, .
\ee

We define $A,B,C$ by 
\be 
A \equiv \,  ({\bf h_1}|{\bf h_1}) \,  , \ \ \ B \equiv   ({\bf h_2}|{\bf h_2})
\, , \ \ \ C \equiv  ({\bf h_1}|{\bf h_2})  \, .
\ee
(Note that the A,B,C terms defined here are, in the single-detector case, 
larger than the A,B,C terms in JKS by a factor of the observation time
$T_0$.)

Then $\Gamma_{11} = \frac{1}{2} A$,  $\Gamma_{22} = \frac{1}{2} B$, and $\Gamma_{12} = \frac{1}{2} C.$  So $\Gamma^{-1}$ takes the form
\bea
\Gamma^{-1} = \frac{2}{D} \left( \begin{array}{cccc}
B & -C & 0 & 0\\
-C & A & 0 & 0\\
0 & 0 & B & -C\\
0 & 0 & -C & A
\end{array} \right).
\label{C}      
\eea
\noindent where $D \equiv AB - C^2$. Thus we arrive at

\bea\label{2F} 
2{\mathcal F} 
= \frac{2}{D}\big[&& \negthinspace B \{({\bf x} | {\bf h_1})({\bf x} | {\bf h_1}) + ({\bf x} | {\bf h_3})({\bf x} | {\bf h_3})\}\nonumber \\
&+& A\, \{({\bf x} | {\bf h_2})({\bf x} | {\bf h_2}) + ({\bf x} | {\bf h_4})({\bf x} | {\bf h_4}) \} \nonumber \\
&-& 2C\, \{ ({\bf x} | {\bf h_1})({\bf x} | {\bf h_2})  + ({\bf x} | {\bf h_3})({\bf x} | {\bf h_4}) \}  
\,\big] \, . \ \ \
\eea

\noindent As a check, consider the case of 
N identical, nearby detectors (assumed to have uncorrelated noises). Then 
$A$, $B$ and $C$ all scale like $N$, while $D \propto N^{-2}$.
In the absence of a GW signal, the only terms in the (implied) double 
sum over $\alpha, \beta $ in (\ref{2F}) that contribute, 
on average, are those with $\beta = \alpha$.
Thus terms like $({\bf x} | {\bf h_1})({\bf x} | {\bf h_1})$ scale 
like $N$ in the absence of a true GW, and
so $<2{\mathcal F}>$ 
remains invariant (always equalling $4$) under changes of $N$ when
there is no true signal. However when there is a true signal, then 
terms like  $({\bf x} | {\bf h_1})({\bf x} | {\bf h_1})$  scale like $N^2$, 
so the non-centrality
parameter $\rho^2$ of the distribution scales like $N$--just as one would
expect.

Eq.~(\ref{2F}) 
can be re-written more compactly if we use complexified
variables, as done in JKS.  Defining
\be\label{FaFb}
2 F_a \equiv ({\bf x} | {\bf h_1} - i {\bf h_3}) \  , \ \ 
2 F_b \equiv ({\bf x} | {\bf h_2} - i {\bf h_4}) \ ,
\ee

Eq.~(\ref{2F}) becomes
\be\label{complex}
2{\mathcal F} = 
\frac{8}{D}\big[\negthinspace B|F_a|^2 + A|F_b|^2 - 
2C \Re (F_a F^*_b)\big] \, .
\ee

\section{${\mathcal F}$-statistic for Multiple Sources}\label{collection}
In this section we consider a search for a {\it collection} of 
$M$ 
nearly periodic GW sources, all with known positions 
and frequencies.  In this case, the signal $h^{\alpha}(t)$ depends linearly
on $4 M$ unknown parameters. Assuming that the M different GW  
frequencies $f_i$ ($i = 1, \cdots , M$) are all sufficiently 
well separated that the detector noises
are uncorrelated (i.e.,$ <{\tilde n}^{\alpha}(f_i) \, {\tilde n}^{\beta}(f_j)^*> = 0$ for $i \ne j$), then a trivial repetition of the arguments in 
\S III shows that the optimum statistic (for either the single-detector
or the multi-detector case) $2{\mathcal F}$  is simply the sum of the  
optimal statistics for the individual sources:
\be\label{sum}
2{\mathcal F} 
\equiv  \sum_i 2{\mathcal F_i} \, .
\ee 
It also easy to show that $y \equiv 2{\mathcal F}$ follows a $\chi^2$ distribution with $4M$ degrees of freedom:
\be\label{pym}
P(y) = \chi^2(y |4M; \rho^2) \, .
\ee
where the non-centrality parameter $\rho^2 = \sum_i \rho^2_i$.

There are currently $\sim 100$
known millisecond (ms) 
pulsars~\footnote{However $\sim 60\%$ of these
are in globular clusters, at distances of several kpc, and so are roughly an
order of magnitude further away than the closest known sources.} 
(defined as pulsars with period $P < 10\,$msec), of which
$\sim 60$ are in binaries. 
We can of course consider any subset of these as some
collection, sum their individual ${\mathcal F}$-statistics
(derived from existing GW data) as in Eq.~(\ref{sum}), and test whether 
or not the collection has been detected. But when is such a 
strategy likely to be advantageous, and for which subsets?
It seems that there are at least two interesting applications of this idea.
First, one might hope that the nearest $\sim 5-50$ (say)
ms pulsars, searched for as a collection, might be more detectable
than any individual member. If this were the case, a multi-source search
might hasten the first discovery of GWs from rotating
neutron stars. 
We shall see below, however, that it is
highly unlikely that a collection of more than a few ($\sim 2-5$)
of the brightest (in GWs) ms pulsars is more detectable than the very brightest
source alone. While it is reasonably likely that the 
brightest few sources, taken together,  
are more detectable than the single brightest one--and we
give a realistic example of this in Sec.IV.D--this will certainly not
be the case for the brightest $20$ or $50$ sources. If there are too many
sources, the strongest ones are effectively diluted by mixing them
with the weaker ones, in the multi-source ${\mathcal F}$-statistic.

To understand the second interesting application of multi-source searches, 
imagine a day when GWs have already been detected from 
the few brightest, closest GW pulsars 
(all at distances of $\sim 0.1 - 0.5\,$ kpc, say), 
but when the GW pulsars in the range $d > 0.5\,$ kpc are still 
too faint to be detected. 
In that situation, it could make sense to take as a collection 
all (or some promising-looking fraction of) the ms 
pulsars in some {\it annulus}--say those in the range $0.5 < d < 1.0\,$ kpc.
This might allow one to measure the {\it average} GW strength of those
more distant sources, even if no single one of them could be positively
detected in GWs, and therefore to begin to make interesting  
statistical statements based on this larger sample. 

We investigate the likely advantages of multi-source searches in the
next five subsections. First, in Sec.IV.A, we consider a 
collection of $M$ sources, for large $M$, and ask: when does adding one 
more source to the collection increase that collection's 
overall detectability?  In Sec.IV.B we re-derive the distribution function 
of signal-to-noise-squared for any spatially uniform population of sources. 
These results from IV.A and IV.B are 
utilized in IV.C, where we 
show that for a uniform planar distribution of GW pulsars 
(representing a somewhat idealized version of the population 
in our neighborhood of the Galactic disk), one might reasonably expect
the few brightest sources, taken together, to be more detectable than
the very brightest one. However this is hardly guaranteed, and any 
advantages of a  multi-source search are likely to be small in this case.
This is illustrated in IV.D, where we consider a fairly realistic example
based on the closet known ms pulsars.  In IV.E we consider  
searching collectively for more numerous, more distant GW pulsars, after
the nearest, brightest ones have been detected, and the advantages
of multi-source searching are shown to be much greater in that case.

\subsection{The large-M case}

Here we compare the sensitivities of a single-source search and a 
search for a collection with M members, when M is much larger than one.
For the single-source
search, the threshold value of ${\mathcal F}$ that gives a $1\%$ FA rate 
is given
by $2{\mathcal F}_{th} (\equiv y_{th}) = 13.277$
(i.e.,$\int_{13.277}^{\infty}\chi^2(y|4) dy = 0.01$).  To be
detectable with FD rate $\ge 50\%$, the signal strength must be at
least $\rho^2 = 10.234$ (i.e.,$\int_{13.277}^{\infty}\chi^2(y|4;
10.234) dy = 0.50$).

By comparison, when M is large, the $\chi^2$ distribution 
with $4M$ d.o.f. is well-approximated by a Gaussian.
Let $y \equiv 2{\mathcal F}$, and let $\rho_{tot}^2 \equiv \sum_{i=1}^{M}\rho_{i}^2 $.
Then 
\be
P(y) = \chi^2(y|4M;\rho_{tot}^2) \approx (8\pi\, M)^{-1/2} e^{-(y - <y>)^2/(8M)}
\ee
\noindent where $<y> = 4M + \rho_{tot}^2 $.
The threshhold value $y_{th}$ such that $\int_{y_{th}}^{\infty} = 0.01$ is then
\be\label{t1}
y_{th} \approx  4M + 4.652 \sqrt{M}  \ \  {\rm (large \,  M)}  \, .
\ee
\noindent (Note that the approximate threshold value that one obtains by
inserting $M=1$ into Eq.~(\ref{t1}) is only 8.652, which
is considerably less than the actual threshold $y_{th} = 13.277$ for the $M=1$ case.
Clearly, this is because the $\chi^2$ distribution with only $4$ d.o.f.
has a substantial tail--i.e., is more skewed to the right than
the higher-$M$ distributions.)

When will a collection be more detectable than its single brightest
member? To answer this, 
let us order the pulsars in the sample such that
\be\label{order}
\rho_1 \ge \rho_2 \ge \cdots \ge \rho_{M}
\ee
Let $T_1$ be the integration time necessary to detect the brightest source, 
and let $T_{coll}$ 
be the integration time required to detect the M-member collection.
For large M, the ratio of these two times is 
\be \label{tpop}
T_{coll}/T_1 
=  0.455 M^{1/2}\frac{\rho^2_1}{\rho^2_{\,tot}}  = 0.455 M^{-1/2}\frac{\rho^2_1}{\rho^2_{\,ave}} \ ,
\ee
\noindent where $\rho^2_{\,ave} \equiv \rho^2_{\,tot}/M$.  
As an extreme example, if $M=25$ and all members have the same
strength ($\rho_1 = \rho_2 = \cdots = \rho_{25})$, then 
$T_{coll}/T_1 = 1/11.0$. (More realistic cases will be considered in
the next two subsections.) 
More generally, we say that a collection is more detectable than
its brightest member if
$T_{coll}/T_1 < 1$. 

How many pulsars should one include in the sample, assuming the
goal is to hasten its detection?
Imagine that pulsars $1, 2, \cdots , M-1$ are included in the
sample, and we want to decide whether to include pulsar $M$.
By Eq.~(\ref{tpop}), the change $\Delta T_{coll}$ in the time required to
confidently detect the collection is
\be
\frac{\Delta T_{coll}}{T_{coll}} = M^{-1}\big(0.5 - \rho^2_{M}/\rho^2_{ave}\big) \, 
\ee
\noindent Thus it is advantageous to increase the sample size 
(because $\Delta T_{coll} < 0$) iff 
$\rho^2_{M} > 0.5 \rho^2_{ave}$. 

Of course, a priori both 
$\rho^2_{M}$ and $\rho^2_{ave}$ are unknown; nevertheless one 
can use both some general statistical arguments and the measured
parameters of nearby millisecond pulsars to make 
a reasonably informed choice. 
We shall illustrate this in the next two subsections.

\subsection{Distribution of $\rho^2$ for Galactic GW pulsars}
What is the distribution function of $\rho^2$ for the
GW pulsars in the Galactic disk, within a few kpc of us?
We can get quite far in answering this question, based on quite
general considerations.

Let $r$ represent a pulsar's distance from the Earth.
For simplicity, we shall consider two different spatial distributions:
a uniform (i.e., homogeneous and isotropic) three-dimensional distribution and a uniform planar distribution.
(These roughly represent the pulsar distributions at distances $r \alt 300\,$pc
and $300\,{\rm pc} \alt r \alt 5 {\rm kpc}$, respectively.)
Let $\sigma(r,f,A,\alpha_i)$ represent the probability density of GW pulsars
in parameter space. Here $f$ is again the NS's gravitational-wave
frequency, $A$ represents the signal's source's intrinsic amplitude
(proportional to $\sqrt{\dot E_{GW}/f^2}$, where $\dot E_{GW}$ is the 
source's GW luminosity), and
the $\alpha_i$ are the relevant angles in the problem. For a
3-D distribution there are $4$ such angles: two for the NS's
angular location on the sky and two for the direction of its spin. For
the planar (2-D) distribution there are only $3$ relevant angles, 
since one angle suffices
for the sky location. For either uniform distribution, the $r-$dependence
can clearly be factored out:
\be\label{sighat}
\sigma(r,f,A,\alpha_i) \equiv F(r)\hat\sigma(f,A,\alpha_i)\, ,
\ee
where 
\bea\label{F}
F(r) &=& 
\left\{
\begin{array}{ll}
4\pi r^2                &\ \  {\rm for \ 3-D} ,      \\
2 \pi H r      &\ \   {\rm for \  2-D} \, .
\end{array}
\right.\
\end{eqnarray}
Here $H \approx 600\,$pc is the thickness of the Galactic disk.

The source's signal-to-noise-squared, $\rho^2$, 
can clearly be written in the following form:
\be\label{rho2}
\rho^2 = A^2 r^{-2} \lambda(f,\alpha_i) \, ,
\ee
where $\lambda(f,\alpha_i)$ is some function of $f$ and the source's angular
parameters.
For a single detector with time-invariant noise characteristics, the 
$f-$dependence can also be factored out of $\lambda(f,\alpha_i)$:
\be\label{lambabar}
\lambda(f,\alpha_i) = \hat\lambda(\alpha_i)/S_h(f) \, ;
\ee
however this factorization of the f-dependence is not necessary 
for our argument.

For notational convenience, we again define $y \equiv \rho^2$.
We now change variables: $(r,f,A,\alpha_i) \rightarrow (y,f,A,\alpha_i)$.
The density function $\sigma$ on the new variables is
\be\label{newsig}
\sigma(y,f,A,\alpha_i) = \sigma(r,f,A,\alpha_i)
\bigg|\frac{\partial(r,f,A,\alpha_i)}{\partial(y,f,A,\alpha_i)}\bigg| \, ,
\ee
where the second term on the rhs of Eq.~(\ref{newsig}) is the Jacobian
of the transformation. It is easy to check that this Jacobian factor is just
\be\label{jacobian}
\bigg|\frac{\partial(r,f,A,\alpha_i)}{\partial(y,f,A,\alpha_i)}\bigg|
 = \frac{1}{2} y^{-3/2} A \lambda^{1/2} \, .
\ee
Combining Eqs.~(\ref{sighat}), (\ref{F}), (\ref{newsig}) 
and  (\ref{jacobian}), 
we therefore have 
\bea\label{newsig2}
\sigma(y,f,A,\alpha_i)&=& 
\hat\sigma(f,A,\alpha_i)\times
\left\{
\begin{array}{ll}
y^{-5/2} \frac{2\pi}{A \lambda^{1/2}}\, ,   & {\rm (3-D)}       \\
y^{-2} \pi H\, ,  &  {\rm \ (2-D)\, . }  
\end{array}
\right.\
\eea
Integrating Eq.~(\ref{newsig2}) over the variables $(f,A,\alpha_i)$, we 
arrive at the  density function for $y$ alone:
\bea
\sigma(y) &=& n_3\, y^{-5/2}  \ \ \ \ {\rm (3-D)} \label{scale3}\\
\sigma(y) &=& n_2\, y^{-2}\, \ \ \ \ \ {\rm (2-D)} \label{scale2}\, ,
\eea
for some constants $n_3$ and $n_2$. 
  
We emphasize that {\it no assumption about the distribution 
of GW pulsars in $f$ and $A$ went into this result}. All that was
required was spatial uniformity--that the nearby pulsars are drawn
from the same distribution as the more distant ones, and that the total
number within some radius scales as $r$ to some power. 
Indeed, the same scaling applies to any source-type having a 
spatially uniform distribution in Euclidean space; 
e.g.,to the extent that one can ignore cosmological effects, the
scaling in Eq.~(\ref{scale3}) also applies to detections of 
binary inspirals.
(To appreciate why this
is at least a  bit remarkable, consider trying to estimate
the density function $\sigma(y)$ for all the pulsars in some 
globular cluster
(say, 47 Tuc). In this case, all the pulsars are effectively at the same 
distance, but we would need to somehow estimate the distribution of GW pulsars
in $f$ and $A$, and then to fold in the detector's noise curve, in order
to estimate $\sigma(y)$ for that cluster.)  
Of course, the scaling laws Eqs.~(\ref{scale3}) is 
well known in other areas of astronomy, where it is the basis 
for the ubiquitous log N-log S test of source strength distributions.

\subsection{Implications of $\sigma(y)$ for collection searches}
We can now return to the question of when a large sample of
the brightest (in GWs) pulsars, taken together, might be more detectable than
the single brightest member. From Eq.~(\ref{tpop}), 
the ``figure of merit'' that
characterizes the detectability of the collection is $M^{-1/2} \sum_{i=1}^{M}
y^i$. In the next two subsections we show how this quantity varies
with collection size for spherical and planar distributions, respectively.

\subsubsection{Uniform 3-D pulsar distribution}
The spherically symmetric case is the less interesting one, from a practical 
standpoint, since there only a few known ms pulsars with $\sim 300\,$ pc
of the Earth. Nevertheless we begin with this case since it 
is somewhat simpler and illustrates our general line of reasoning. 

Imagine that we have included in our collection all the brightest
(in GWs) ms pulsars, down to some lower limit $y_{\tt l}$. Then
$M \approx \int_{y_{\tt l}}^{\infty}  n_3\, y^{-5/2}$ and  
\be
\sum_{i=1}^{M} y^i \approx \int_{y_{\tt l}}^{\infty}  n_3\, y^{-3/2} \, dy \, ,
\ee
so 
\be
M^{-1/2} \sum_{i=1}^{M}y^i  \approx \sqrt{6 n_3}\  y_{\tt l}^{1/4}
\ee
This is a strictly increasing function of  $y_{\tt l}$ (albeit that 
it increases rather slowly).
But increasing $y_{\tt l}$ means shrinking the collection.
Thus for a uniform 3-D distribution, it would be unlikely that
a large collection of the brightest sources would be detectable before
the single brightest member was detected.

\subsubsection{Uniform 2-D pulsar distribution}
We turn now to the planar case. We begin by considering the detectability of
all GW pulsars with $\rho^2$ in the interval $y_{\tt l} < \rho^2 < y_{\tt u}$, 
(so $y_{\tt l}$ and $y_{\tt u}$ are the lower and upper limits 
of the interval, respectively).
Again, assuming the number of sources in the interval is large, 
the appropriate 
figure of merit, characterizing the detectability of the
whole collection, is  $M^{-1/2} \sum_{i=1}^{M} y^i$. The continuous version of
this is clearly 
\bea
M^{-1/2} \sum_{i=1}^{M}y^i  &\approx& \bigg[\int_{y_{\tt l}}^{y_{\tt u}}{n_2\, y^{-2}}\, dy\bigg]^{-1/2}    \int_{y_{\tt l}}^{y_{\tt u}}{n_2 y^{-1}}\, dy \nonumber \\
 & = & n_2^{1/2}\,{\rm ln}(y_{\tt u}/y_{\tt l})
\bigg[y_{\tt l}^{-1} - y_{\tt u}^{-1}\bigg]^{-1/2}  \\    
& = & (n_2\, y_{\tt u})^{1/2} \bigg[\frac{-{\rm ln}(x)}{(x^{-1} - 1)^{1/2}}
\bigg]  \label{merit3}
\eea
where in the last line we introduced the dimensionless ratio 
$x \equiv y_{\tt l}/y_{\tt u} < 1$. We gain some insight into Eq.~(\ref{merit3}) if we 
re-express $n_2$ in terms of $y_{max}$, which we define to be
the $y$-value of the strongest Galactic source. Let $\tilde y_{max}$ 
represent the median value of $y_{max}$, for our distribution function 
Eq.~(\ref{scale2}).
Then $\tilde y_{max}$ is given implicitly by 
\be\label{ymax}
\int_{\tilde y_{max}}^{\infty}n_2\, y^{-2}\, dy = 0.5\, ,
\ee
(since then there is a $50\%$ chance of finding a stronger source than 
$\tilde y_{max}$), so $\tilde y_{max} = 2 n_2$. The actual value of 
$y_{max}$ for our Galaxy is therefore $y_{max} = 2 \beta n_2$, where
we expect $\beta$ is some number of order one.
The rhs in Eq.~(\ref{merit3}) can therefore be written as
\be\label{rhsmerit3}
\bigg(\frac{y_{max} y_{\tt u} }{2\beta}\bigg)^{1/2} 
\bigg[\frac{-{\rm ln}\,(x)}{(x^{-1} - 1)^{1/2}}\bigg] \, .
\ee
Next we consider the function 
$f(x) \equiv -{\rm ln}\,x/\big(x^{-1} - 1\big)^{1/2} $, which 
is displayed in Fig.~1. 

\begin{figure}[ht!]
\centerline{\includegraphics[width=8.0cm]{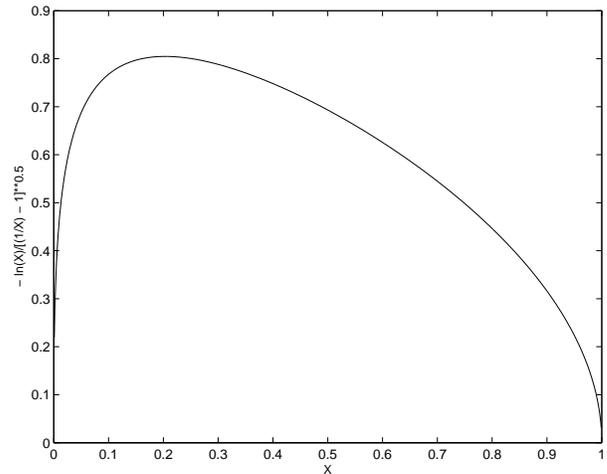}}
\vspace{0mm}
\caption{Plot of function
$f(x) \equiv {-\rm ln}(x)/\big(x^{-1} - 1\big)^{1/2}$, which displays a 
maximum at $x \approx 0.203$.}
 \label{fx}
\end{figure}
Note that it has a maximum at
$x \approx 0.203$, where $f(x) \approx 0.805$.
Thus an optimized source collection has a ratio of weakest-to-brightest  
source (in terms of their signal-to-noise-squared) of $\sim 1/5$. 
However the maximum in $f(x)$ is rather broad; at a brightness ratio 
of $20$ ($x = 0.05$), $f(x)$ has decreased only $\sim 15\%$, to $0.687$.
Assuming the collection includes all the brightest sources down to some 
limiting brightest $y_{\tt l}$, the number of sources in the collection is
\bea
M &\approx& n_2/y_{\tt l} = \tilde y_{max}/(2\,y_{\tt l}) \\ 
&=& 2.5\, \beta^{-1}\, \big(\frac{y_{max}/y_{\tt l}}{5}\big) \, .
\eea
Thus including all sources down to strength $y_{\tt l}$ while also
optimizing $y_{\tt l}/y_{max}$ leads to a rather small value of $M$, 
which is then somewhat 
outside the range of validity 
of the Gaussian approximation that led to our ``figure of merit''
$M^{-1/2}\sum_{i=1}^{M} y^i$; nevertheless it is clear that the
``most-detectable'' collection will have at most a few members.

To estimate $T_{coll}/T_1$ (the time to detect this collection divided by
the time to detect the single brightest source), it would seem we need to
evaluate the integral $\int_{y_{\tt l}}^{\infty}{n_2 y^{-1}}$ 
(i.e., the continuous version of $\sum_{i=1}^{M} y^i$), which is 
logarithmically divergent. Physically, though, it seems sensible 
to simply cut off the upper end of the integral at 
some $y_{cut}$ of order $y_{max}$. I.e., we cut off the integral at the
y-value of the brightest source.~\footnote{Of course, our planar approximation 
breaks down at $r < H/2$, and this ``switchover'' from an effective 
$2-D$ to a $3-D$ distribution at short distances would obviate the
need for an artificial cutoff in a more realistic treatment of this 
problem.}

Thus, setting $y_{\tt u}$ equal to  $y_{max}$ in Eq.~(\ref{rhsmerit3}) and
plugging the result into Eq.~(\ref{tpop}), 
we estimate
\be\label{tpop2} 
\frac{T_{coll}}{T_1} \approx 0.455 (2\beta)^{1/2}/f(0.203) \approx  0.80 \beta^{1/2} \, .\ee
Again, the use of Eq.~(\ref{tpop}) in deriving  Eq.~(\ref{tpop2})
is strictly valid only for large M; nevertheless the basic moral
is clear: $T_{coll}/T_1$ is of order unity, and whether in actual experience
it is greater or less than one depends strongly on $\beta$, i.e., on whether
the strongest source is stronger or weaker than one would expect, based
on the source distribution function.

\subsection{Example: The Best Candidates among Known Millisecond Pulsars}
We next consider a potentially realistic example: a collection drawn from
the population of known millisecond pulsars. 
We attempt to construct the ``most detectable'' collection from these.
Of course, we do not know their
actual GW strengths, so for this exercise we will estimate their
strengths by assuming that they all have the same non-axisymmetry 
$I \epsilon \equiv I_{xx} - I_{yy}$ (where the NS is assumed to be spinning
about its z-axis).  This non-axisymmetry might be generated, e.g., by 
lateral variations in the crustal composition or  
strong toroidal magnetic fields in the NS interiors~\cite{cutler-thorne}.

For each of the millisecond pulsars, we estimate $\rho^2$ as follows. 
First, we estimate $h_0$ at the Earth 
from the pulsar's measured spin
and the
best available estimate of its distance $r$~\cite{atnf}, 
using
\be\label{h0eps}
h_0 = 4\pi^2 (G/c^4)\, I \epsilon \, f^2 \, r^{-1}
\ee
where here we will assume the GW frequency $f$ is exactly twice the
pulsar's measured spin frequency, $\nu$.
Then we estimate $\rho^2$ using
\be\label{rho2}
\rho^2 = 2 \bigg(\frac{h_0^2 \, T_0}{S_h(f)} \bigg) K(\alpha_i)
\ee
where $T_0$ is some fiducial observation time and 
$K(\alpha_i)$ is factor that depends
on the sky-location and spin-orientation of the source.
The spin-orientations of the millisecond 
sources are poorly constrained, so for simplicity, in our estimates, we will 
simply replace K by its average value (over all angles). 
For $S_h(f)$, we use the values for the Advanced LIGO noise curve, as 
generated by the Bench software package~\cite{bench}.
(Eq.~(\ref{rho2}) is for a single detector; if one optimally combined 
the outputs of 
LIGO's L1, H1 and H2 interferometers, 
then $\rho^2$ should be 
approximately $2.25$ times greater than for either L1 or H1 alone.)

Given the above inputs, we find that there are $5$ ms pulsars that stand out
as the best candidates for detection by the Advanced LIGO Interferometers.
They are PSRs J0437-4715, J0030+0451, J2124-3358, J1744-1134, and J1024-0719.
These $5$ are also the closest known ms pulsars.
And pulsar 1 (PSR J0437-4715), which at $d = 0.14\,kpc$ is 
the closest of all the known ms pulsars, is estimated to be the strongest
GW source.
Relative to pulsar 1, the GW strengths of the other four sources
are given by: $(\rho_2/\rho_1)^2 = 0.38$, $(\rho_3/\rho_1)^2 = 0.32$,  
$(\rho_4/\rho_1)^2 = 0.17$, and $(\rho_5/\rho_1)^2 = 0.16$.


Assuming the above estimates of $\rho^2$ for the five best candidates
were correct, what would be $T_{coll}/T_1$ (the ratio of the integration 
times necessary to detect the 5-member collection and the brightest 
individual source)? For our 5-member sample, the threshold value
for detection with $1\%$ FA rate is $y_{th} = 37.57$
(i.e.,$\int_{37.57}^{\infty}\chi^2(y|20) dy = 0.01$).  To be
detectable with FD rate $\ge 50\%$, the signal strength must be at
least $\rho^2_{tot} = 18.45$ (i.e.,$\int_{37.57}^{\infty}\chi^2(y|4;
18.45) dy = 0.50$). Thus
\be\label{ex2}
\frac{T_{coll}}{T_1} = \big(\rho^2_{1}/\rho^2_{tot}\big) 
\big(\frac{18.45}{10.23}\big) = 0.87 \, .
 \ee
E.g., if it took two years to confidently detect the strongest
source, the 3-member ensemble would be detectable in 
about 21 months.  
(Note that in deriving Eq.~(\ref{ex2}) we have used the 
actual $\chi^2$ distribution with $20$ d.o.f., 
not the Gaussian approximation to it.)

Should we add a sixth pulsar to the sample? 
From the analysis in the previous subsection, this would be advantageous
if $\rho^2_6/(\sum_{i=1}^{5}\rho^2_i) > 0.10$.
But we estimate that the sixth most detectable pulsar is J1012+5307, with 
$(\rho_6/\rho_1)^2 = 0.07$, so we restrict the sample to the 
most promising five.
[Indeed, if we had restricted the sample to only the most promising 3 pulsars, 
we would coincidentally have arrived at the same estimate for   
$T_{coll}/T_1$. For the 3-member case,  $(\rho^2_{tot}/\rho^2_1) = 1.70$, while
the threshold for detection with $1\%$ FA rate is $y_{th} = 26.22$, and
the signal strength must be $\rho^2_{tot} \ge 15.13$ to be detectable
with a FD rate $\ge 50\%$. Thus we would estimate
$T_{coll}/T_1 = 15.13/(2.03*10.23) = 0.87$, the same as for the 5-member
collection. However we highlighted the 5-member result since that one is
clearly somewhat more robust against deviations of the actual source
strengths away from our fiducial estimates.]


Clearly, since the actual orientations of the ms pulsars 
are unknown and the distances are known only to within a factor
$\sim 2$, the above estimate merely gives a 
rough indication of the time-savings 
that a multi-source search might reasonably lead to.

We also note that if we were to estimate source strengths by assuming
that all ms pulsars are spinning down primarily due to GW emission, 
thus using
$h_0 = \big(\frac{5 G I \dot\nu}{2 c^3 \nu}\big)^{1/2} r^{-1}$
instead of Eq.~(\ref{h0eps}), and then repeat the above analysis 
from that starting point, we would find that {\it no} subset of the 
known ms pulsars is more
detectable than the single brightest source, PSR J0437-4715.
This just highlights the fact that the ratio 
$T_{coll}/{T_1}$--and especially whether that ratio is
greater or less than one--depends 
rather sensitively on the relative strengths of the few brightest 
sources, which of course we will not know in advance.


\subsection{Search for a collection of weaker GW pulsars}
The last two subsections showed that a search for a collection of the
very brightest GW pulsars may offer some advantages, compared to 
a search for the very brightest one, but any such advantages are likely to
be quite modest. 
We now turn to a case where the advantages of a whole-collection
search are much more impressive. 

Consider some time in the future, when 
the few brightest GW pulsars are presumed to have already been detected. These
are presumably among the closest GW pulsars, while the GWs from their
more distant cousins are still too weak (at the Earth) to be detected. 
Now once again consider collecting together all known ms pulsars in the 
range $y_{\tt l} < y < y_{\tt u}$.  (Of course, again, one does not know
precisely which these are, but whatever lessons are learned from the
brightest GW pulsars, combined with the known distances and spin rates 
of the remaining millisecond pulsars, will probably allow one to
make fairly educated guesses.) If all ms pulsars had the same intrinsic GW
strength, the same frequency, and the same angular factor $K(\alpha_i)$ (from Eq.~\ref{rho2}), then clearly these pulsars would occupy a circular annulus in the
disk.  
In fact, of course, these GW pulsars will {\it not} have the same intrinsic
amplitude, frequency, or angular factor, but we still find it helpful, 
conceptually, to imagine the GW pulsars with $y_{\tt l} < y < y_{\tt u}$ as
filling a roughly annular region.

For an optimally chosen annular region (one that minimizes the integration 
time required for positive detection), what is the optimal
value of $x \equiv y_{\tt l}/y_{\tt u}$. We worked this out in Sec.IV.C.2;
the optimum selection has $x \sim 1/5$.  How many sources are in this
range? For a planar distribution, the answer is clearly
\be
M \approx
\int_{y_{\tt u}}^{y_{\tt u}/5}
n_2\, y^{-2}\, dy = 2\big(\frac{\tilde y_{max}}{y_{\tt u}}\big) \, ,
\ee
where we have used $n_2 = \tilde y_{max}/2$. Similarly the ratio 
$T_{coll}/T_{\tt u}$ 
(where $T_{\tt u}$ is the integration time required
to detect a GW pulsar whose $\rho^2$ equals  $y_{\tt u}$) is given by
\bea
T_{coll}/T_{\tt u} &\approx& 0.455\, y_{\tt u}/[(n_2 y_{\tt u})^{1/2} 
f(x \approx .203)] \\
&\approx& 0.8\, \big(\frac{y_{\tt u}}{\tilde y_{max}}\big)^{1/2} \, .
\eea

For example, if $y_{\tt u}/\tilde y_{max} = 1/10$, then the optimal
number of sources for the ``annulus'' is $M \approx 20$, and the
integration time required to detect that whole collection of 
$20$ GW pulsars is only $0.8/\sqrt{10} \approx 0.25$ as long as the 
time required to detect the brightest single member in that group.
Again, such a detection would provide an estimate of the average $\rho^2$ 
for ms pulsars in that collection, even though none could be detected
individually in GWs.

\section{Conclusions}
\label{conclusion}

The most sensitive GW detectors 
(currently the LIGO L1, H1, and H2 
interferometers) have very similar sensitivities, and this is likely to 
remain the case for some years. 
In such a case, one
can significantly increase the effective signal-to-noise of 
any source by optimally combining the data streams.
Here we have derived the appropriate formulae for doing so, for
GW pulsar searches.
For $N$ GW detectors with the same sensitivity, the observation 
time $T_{det}$ required to detect any particular GW pulsar 
scales like $N^{-1}$, and so combining the data streams this way
is clearly a useful strategy.

However we remind the reader that 
our analysis of the multi-detector statistics has
assumed the noise is Gaussian. In the more realistic case, one 
would one want to veto candidate 
detections that had a large ${\mathcal F}$-statistic 
but that did not sufficiently resemble actual GW pulsar signals, 
e.g., because 
the relative sizes of the signal in the various detectors did 
not conform with expectations for {\it any} choice of parameters
$(\lambda^1,\lambda^2,\lambda^3,\lambda^4)$.  In particular, we 
imagine that a realtistic implementation would incorporate some
multi-detector version of the chi-square veto developed in 
Itoh et al.~\cite{Itoh}; however we have not considered this in
any detail.

We have also pointed out that one can search for collections of pulsars, 
and that the optimal frequentist search for such
collections simply adds up the ${\mathcal F}$-statistics
of the individual members. We considered two cases in detail. 
We first asked whether the few brightest GW pulsars might be
discovered, collectively, before the very brightest one.
The answer turns out to depend rather sensitively on the relative strengths of
the few brightest sources, and so we can only equivocate:
maybe yes, maybe no. But even if some collection turns out to be
more detectable than the single brightest source, it is unlikely to
``win'' by much. However, {\it after} the few brightest GW pulsars have
been discovered, searching for more distant pulsars by summing their
${\mathcal F}$-statistics should prove to be an effective strategy, allowing
one to measure the average strength of many sources that are not
individually detectable.

\begin{acknowledgments}
We thank Graham Woan for very helpful correspondence and thank the
anonymous referee for helpful commens. C.C.'s work was
partly supported by NASA Grant NAG5-12834.
\end{acknowledgments}

\end{document}